\documentclass[AMA,STIX1COL]{WileyNJD-v2}

\articletype{Article Type}%

\received{01 June 2018} 
\revised{01 August 2018}
\accepted{01 August 2018}

\usepackage{subcaption}
\usepackage{matlab-prettifier}
\usepackage{amsmath, amsfonts, amssymb}

\usepackage{glossaries}
\makeglossaries

\begin{document}

\title{Open-source platforms for fast room acoustic simulations in complex structures}

\author[1]{Matthieu Aussal}

\author[2]{Robin Gueguen}

\authormark{AUSSAL \& GUEGUEN}

\address[1]{\orgdiv{Centre de math\'ematique appliqu\'ees}, \orgname{\'Ecole Polytechnique}, \orgaddress{91128 Palaiseau, \country{France}}}

\address[2]{\orgdiv{Institut des Sciences du Calcul et des Donn\'ees}, \orgname{Sorbonne Universit\'e}, \orgaddress{Campus Pierre et Marie Curie - 4 place Jussieu, 75252 Paris Cedex 05 \country{France}}}

\corres{ \email{matthieu.aussal@polytechnique.edu}\\\email{gueguen.robin@gmail.com}}

\abstract[Summary]{
This article presents new numerical simulation tools, respectively developed in {\sc  Matlab} and \textit{Blender} softwares. Available in open-source under the GPL 3.0 license, it uses a ray-tracing/image-sources hybrid method to calculate the room acoustics for large meshes. Performances are optimized to solve problems of significant size (typically more than 100,000 surface elements and about a million of rays). For this purpose, a \textit{Divide and Conquer} approach with a recursive binary tree structure has been implemented to reduce the quadratic complexity of the ray/element interactions to near-linear. Thus, execution times are less sensitive to the mesh density, which allows simulations of complex geometry. After ray propagation, a hybrid method leads to image-sources, which can be visually analyzed to localize sound map. Finally, impulse responses are constructed from the image-sources and FIR filters are proposed natively over 8 octave bands, taking into account material absorption properties and propagation medium. This algorithm is validated by comparisons with theoretical test cases. Furthermore, an example on a quite complex case, namely the ancient theater of Orange is presented. 
}


\jnlcitation{\cname{%
\author{M. Aussal}, and
\author{R. Gueguen}, 
} (\cyear{2018}), 
\ctitle{Open-source platforms for fast room acoustic simulations in complex structures}, \cjournal{}, \cvol{2018;00:1--14}.} 

\maketitle

\section*{Introduction}
\label{sec1}
\begin{figure}[t]
\centering
	\includegraphics[width=0.5\linewidth]{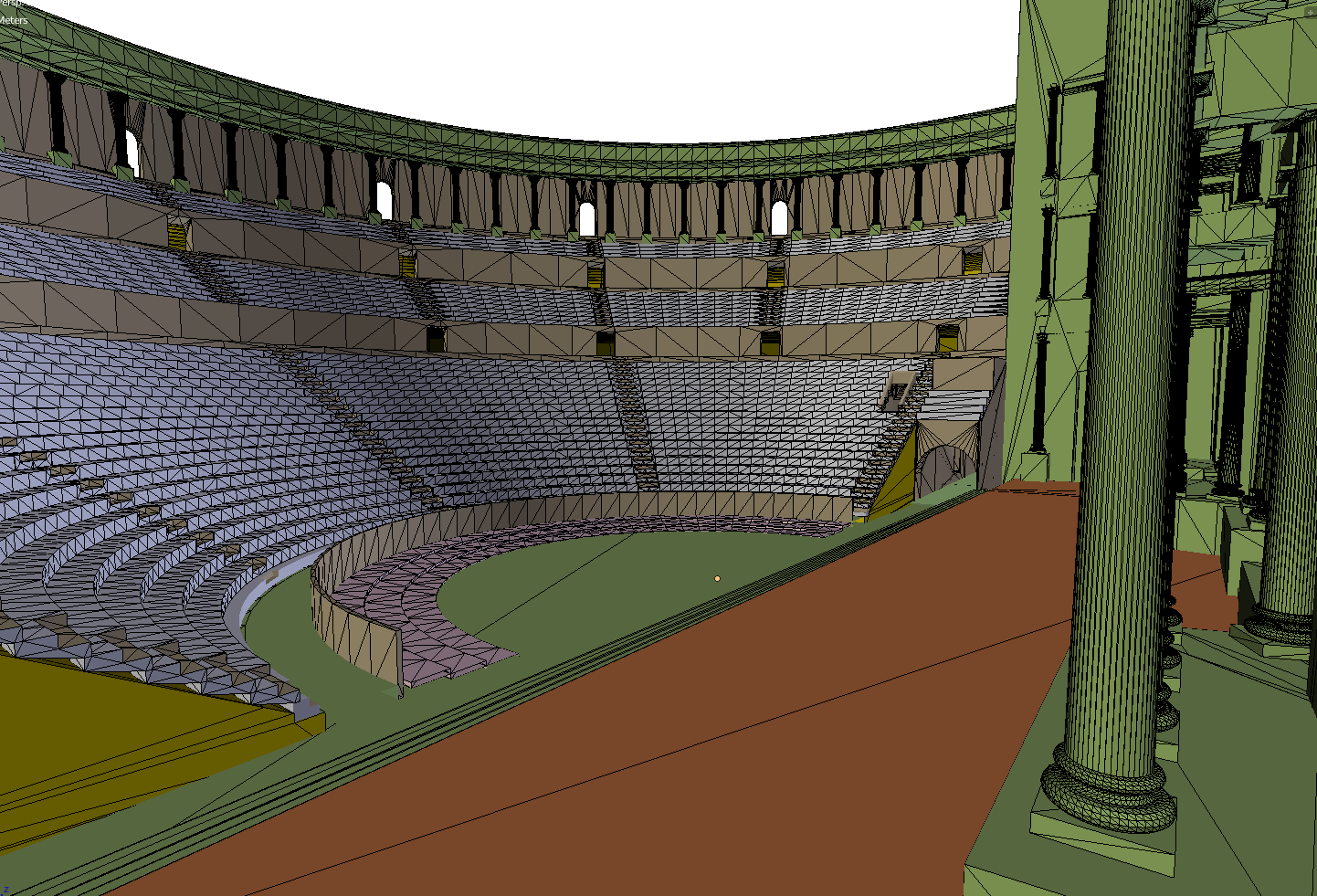}
	\caption{Mesh of restituted theater of Orange modelized on \textit{Blender} (436~000 triangles).}
	\label{maillage}
\end{figure}

Today, digital technologies allow research to explore previously inaccessible areas, as virtual reality for archaeology. In this domain, many works focus on the visual restitution, but acoustic studies can reinforce researches to improve the understanding of the ancient world. For example, during the Roman Empire, architects have designed buildings using acoustic rules  \cite{vitruve}. In this study, we focus on the ancient theater of Orange which has a significant size (100m wide), a complex geometry (ornaments, bleachers, columns, arches, etc.) and which is open-air. In a previous work, a complete mesh was designed using \textit{Blender} CAD software \cite{doc_blender}. To be representative, this mesh, shown in the figure \ref{maillage}, processes 436~000 elements (triangular faces) and represents the actual archeological knowledge precisely \cite{theseRobin}. As the mesh size makes difficult the use of precise methods (FEM, BEM, etc.) \cite{gypsilab}, ray-tracing approximation was performed, in order to compute fastly the full-band room impulse response (50 to 15000Hz). Under this assumption, open-sources platforms were developed following two steps. We first build a prototype using \textit{Gypsilab}, an open source {\sc Matlab} framework for fast prototyping \cite{gypsilab}. This preliminary work was useful to construct and validate ideas and algorithms. It leads to the creation of a new toolbox, \textit{openRay}, now appended to the master branch of \textit{Gypsilab} and freely downloadable \cite{githubGypsi}. In a second step, all algorithms were retranscrypted in C++ using \textit{Qt Creator}, leading to an autonomous tool, \textit{Just4RIR}. A python interface was added, in order to use this library as a \textit{Blender} plug'in. At the end, this plug'in allow archeologists to only work on \textit{Blender}, modifying easily meshes and materials, run acoustic simulation and visualize results. \\
After reminders on acoustical energy propagation represented by ray-tracing, this paper gives implementation details of the method that was used to obtain a fast computation for large meshes. At the end, validation test cases are given and application on a virtual model of the ancient theater of Orange is performed. 

\section{Acoustical energy modelization}\label{sec2}
\subsection{Continuous domain equation}

By modelling a point sound source located at 0 as a localized pulse in space, the associated acoustical energy E(t) propagates \cite{jouhaneau} over time on a spherical surface $S(t)$ centered in 0, as :
\begin{equation} 
E(t) = E_0 \int_{S(t)} \overrightarrow{I}(t)\cdot\overrightarrow{ds} \qquad \forall t > 0,
\end{equation}
with $E_0$ the initial energy and $\overrightarrow{I}(t)$ the acoustical intensity. According to the first principle of thermodynamics and by neglecting the effects of losses related to the absorption of the propagation medium, the acoustic energy is preserved over time and we may normalize the source in such a way that :
\begin{equation} 
\int_{S(t)} \overrightarrow{I}(t)\cdot\overrightarrow{ds} = 1 \qquad \forall t > 0.
\label{eq_2}
\end{equation}
The propagation being isotropic, we deduce that :
\begin{align} 
|| \overrightarrow{I}(t) || &= \frac{1}{4\pi d(t)^2} \qquad \forall t > 0,
\end{align}
reflecting that intensity decreases as the square of the distance to the source $d(t)$. The energy carried by a solid angle $\Omega_{\sigma}$ is obtained by integrating on the portion $\sigma(t)$ of $S(t)$ and satisfies :
\begin{equation}
E_{\sigma}(t) = E_0 \int_{\sigma(t)}  \frac{1}{4\pi  d(t)^2} ds = \frac{E_0}{4\pi}  \Omega_{\sigma}.
\label{eq_4}
\end{equation}
The energy of a solid angle is constant over time and corresponds to a portion of the initial energy~$E_0$. Thus, subdividing $S(t)$ in $N$ portions $\sigma_i(t)$, the total energy can be decomposed as a sum of elementary energies, carried by corresponding solid angles $\Omega_i$, such as : 
\begin{equation}
E(t) = \sum_{i=1}^N E_i(t) = \frac{E_0}{4\pi}  \sum_{i=1}^N \Omega_i  \qquad \forall t > 0.
\label{eq_5}
\end{equation}
One can notice that $(\Omega_i)_{i\in[1,N] }$ is a directional basis, representing the energy propagation by piecewise constant elements. Furthermore, for greater clarity, we definitively set in the following $E_0 = 1$.

\subsection{Discrete model}

\begin{figure}[t]
\centering
	\includegraphics[width=0.6\linewidth]{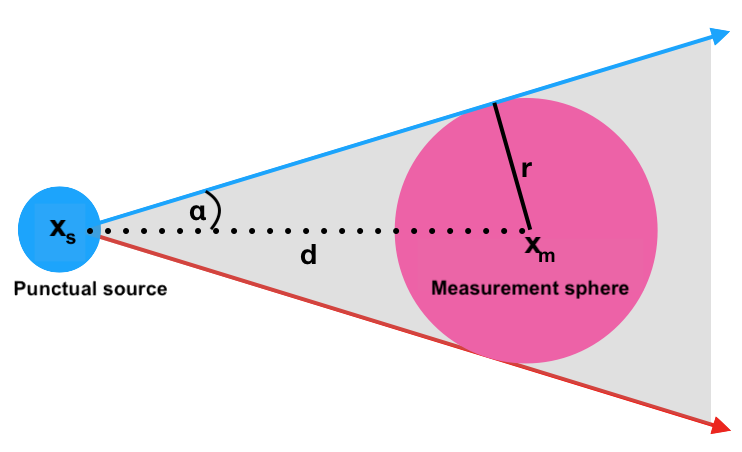}
	\caption{Representation of a $r$-radius measurement sphere centered in $x_m$, receiving energy from a sound source in $x_s$.}
	\label{schema_rayon}
\end{figure}

To numerically represent the energy propagation, we have to discretize basis $(\Omega_i)_{i\in[1,N] }$ in equation \ref{eq_5}. For this purpose, we define a ray object composed by :
\begin{itemize}
\item Its origin $x_i$,
\item Its direction vector $\overrightarrow{u_i}$,
\item The energy that it carries $E_i$.
\end{itemize}

For example, with an omnidirectional source\footnote{For a directional source, a non-uniform spatial sampling may be used.} considered before, $N$ rays are given by :
\begin{itemize}
\item The source coordinate ($x_i = x_s,~\forall i\in[1,N]$),
\item A unit sphere uniform sampling (e.g.~icosahedre subdivision, Fibonacci's rule \cite{fibonacci}, etc.),
\item An uniform energy repartition ($E_i = \frac{4\pi}{N},~\forall i\in[1,N]$).
\end{itemize}

\begin{figure}[t]
\centering
	\includegraphics[width=0.65\linewidth]{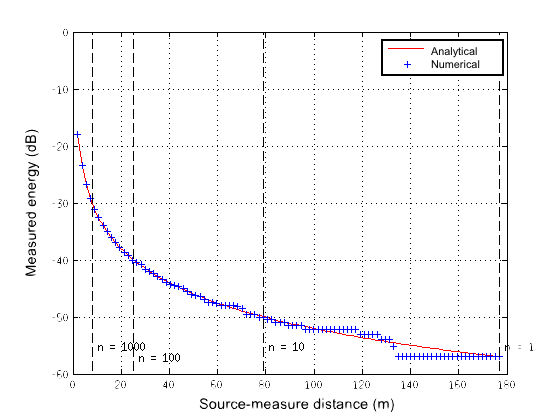}
	\caption{Measured energy (dB) in function of distance between $x_s$ and $x_m$ in meter for $r = 0.36$m and $N = 10^6$. Blue crosses stand for the statistical measure $f(r) = \frac{n(r)}{N}$ and red ligne the analytic function $f(r) = \frac{\pi r^2}{4\pi d^2}$. (computed on \textit{Gypsilab})}
	\label{energie}
\end{figure}

To complete this approach, we have to define a discrete measure of energy propagation. To this end, we consider a $r$-radius measurement sphere $S(x_m, r)$, centered on $x_m$ (fig~\ref{schema_rayon}).  We can then add the contributions of a $n$-rays  beam that intersect this sphere to calculate the acoustic energy $E_m$ at the point $x_m$ :

\begin{equation}
E_m \approx  \frac{1}{4\pi}  \sum_{i=1}^n E_i.
\end{equation}
In the particular case of an omnidirectionnal source, we have : 
\begin{equation}
E_m \approx  \frac{n}{N},
\label{eq_7}
\end{equation}
which means that the measured energy $E_m$ is statistically and naturally represented by the ratio between the number of rays forming a beam to the total number of rays. This formula is nothing but the discretization of the continuous model (eq.~\ref{eq_4}) in which the measured energy is given by :
\begin{equation}
E_m = \frac{1}{4\pi}  \Omega_m,
\end{equation}
where $\Omega_m$ is a solid angle at which the measurement sphere is seen from $x_s$. Using the notation of the figure \ref{schema_rayon}, we have :
\begin{equation}
\Omega_m = 2\pi(1-\cos{\alpha}) = 2\pi \left( 1 - \sqrt{1-\frac{r^2}{d^2}} \right).
\end{equation}
Considering $\frac{r}{d} \ll 1$, we observe that
\begin{equation}
\Omega_m \approx \pi \frac{r^2}{d^2}.
\end{equation}
which entails
\begin{equation}
E_m \approx  \frac{n}{N} \approx  \frac{\pi r^2}{4\pi d^2}.
\label{eq_12}
\end{equation}
To ensure the existence of this last approximation, beam has to be measurable and count at least one ray ($n\geq1$). This assumption is crucial to ensure the validity of the concept. Thus, fixing a measurement radius r, approximation (\ref{eq_12}) gives a maximum range of the discrete model :  
\begin{equation}
\label{eq_13}
	d \leq \frac{r}{2}\sqrt{\frac{N}{n}}.
\end{equation}
In addition, figure \ref{energie} shows how this modelization fills with distance between source and measures. The accuracy of the measurement depends strongly on the number of  rays counted, then, the more $n$ increases, the more accurate will be the measurement. Nevertheless, in practice, values for a short distance between the source and the measurement sphere represent direct sound and first reflections, whereas long distances describe the diffuse field. Under this assumption, we can consider this model acceptable for all beam such as $n\geq1$. 

\begin{figure}[t]
	\centering
	\includegraphics[width=0.5\linewidth]{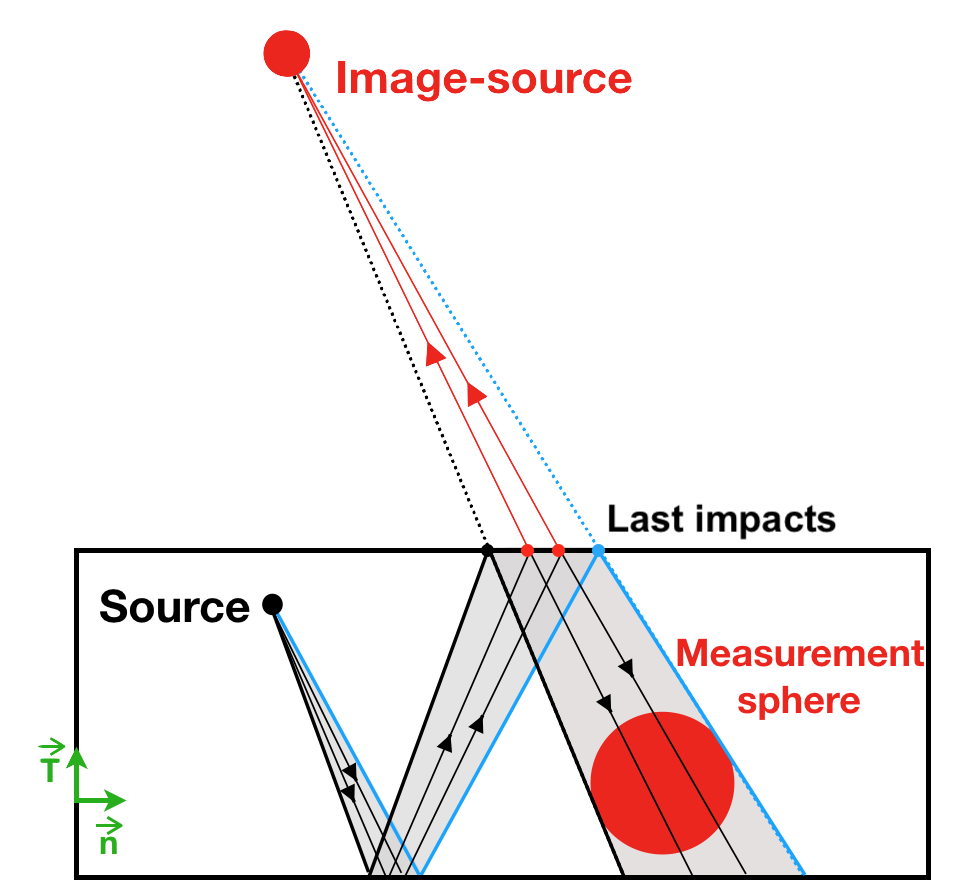}
	\caption{Sketch of the creation of an image-source by successive reflections of a ray on the walls of a room.}
	\label{schema_SI}
\end{figure}

\subsection{Presence of an obstacle}
For the case of acoustic propagation in the presence of an obstacle, we choose to consider only specular reflections (Snell-Descartes laws). Indeed, this approximation is suitable when surfaces are large in comparaison to wavelengths, because diffraction effects can be neglected \cite{jouhaneau}. For a room, this condition is reached if :
\begin{equation}
ka \gg 1, 
\end{equation}
with $k$ the wave number and $a$ the characteristical diameter of the room \cite{hautes_freq}. This approach is currently used by room acoustic softwares (e.g. \textit{Odeon} \cite{odeon}, \textit{Grasshopper} \cite{grasshopper}, etc.) regarding to audible frequency range (62,5 to 15000Hz). In particular, as the theater of Orange has a characteristical diameter of about 50 meters, the high frequency approximation is clearly valid. 

Following the discrete model, when an incident ray intersects a flat surface, a reflected ray is generated from the collision point. Noting $\overrightarrow{u_i}$ the direction vector of the incident  ray, the reflected direction vector $\overrightarrow{u_r}$ is defined by :
\begin{equation}
\label{eq_15}
\overrightarrow{u_r} = (\overrightarrow{u_i} \cdot \overrightarrow{T})\overrightarrow{T} - (\overrightarrow{u_i} \cdot \overrightarrow{n})\overrightarrow{n},
\end{equation}
with $\overrightarrow{T}$ the tangent basis and $\overrightarrow{n}$ the normal vector to the surface. Moreover, the energy of the reflected ray is obtained by :
\begin{equation}
E_r(f) = E_i(f)(1 - \alpha(f)),
\end{equation}
with $\alpha(f)$  the absorption coefficient of the surface that depends on of the frequency $f$. Practically, the absorption coefficients are often given per octave bands and can be found in various databases. Both \textit{Gypsilab} and \textit{Just4RIR} use the open access \textit{Odeon} database \cite{odeon} defined on eight octave bands (see table \ref{tab_coeff_abs}). 

Finally, considering wall absorption, energy measured statistically (eq. \ref{eq_12}) is extended by :
\begin{equation}
E_m(f) \approx  \frac{n}{N}(1 - \alpha(f)),
\end{equation}
that we generalize to
\begin{equation}
E_m(f) \approx  \frac{n}{N}\prod_{j=1}^{m}(1 - \alpha_j(f)),
\label{eq_18}
\end{equation}
in the case of $m$ reflexions.

\begin{table}[t]
\centering
	\begin{tabular}{| c | m{2.5cm} | *{8}{c|}}
		\hline
		Reference & Material name & 62,5Hz & 125Hz & 250Hz & 500Hz & 1kHz & 2kHz & 4kHz & 8kHz \\
		  \hline
		  \hline
		   1 & 100\% absorbent & 1 & 1 & 1 & 1 & 1 & 1 & 1 & 1 \\
		   \hline
		2 & 100\% reflecting & 0 & 0 & 0 & 0 & 0 & 0 & 0 & 0 \\
		   \hline
		107 & Concrete block, coarse\footnotemark & 0.36 & 0.36 & 0.44 & 0.31 & 0.29 & 0.39 & 0.25 & 0.25 \\
		   \hline
		3000 & Hollow wooden podium\footnotemark & 0.4 & 0.4 & 0.3 & 0.2 & 0.17 & 0.15 & 0.1 & 0.1 \\
	     \hline
	 \end{tabular}
	\caption{Examples of absorption coefficients given in the online \textit{Odeon} database \cite{odeon}.}
	 \label{tab_coeff_abs}
\end{table}
\addtocounter{footnote}{-1}
\footnotetext{Harris, 1991}
\addtocounter{footnote}{1}
\footnotetext{Dalenback, CATT}

\subsection{Image-sources}
\label{is}
Although the generalized formulation (\ref{eq_18}) may be sufficient to generate room acoustic data, we also construct images-sources from the path of rays. To this end, when rays intersect the measurement sphere and following the reverse return principle, they are retro-propagated along the last direction vector. Thus, from this measurement sphere, rays focus on punctual images-sources (see fig. \ref{schema_SI}). Each image-source is then located relatively to the listener and carries an energy according to equation (\ref{eq_18}). By noting $(x_s)_{s \in [1, N_s]}$ the relative position of the $N_s$ image-sources and $(E_s)_{s \in [1, N_s]}$ the associated energy, couples $(x_s;E_s(f))_{s \in [1, N_s]}$ contain all useful informations for room acoustic analysis and auralization. 

First of all, relative distance of each image-source $(d_s)_{s \in [1, N_s]}$ can be computed. This distance is also used to take into account the air absorption, by modifying equation (\ref{eq_18}) into :
\begin{equation}
E_s(f) \approx  \frac{n}{N}  e^{-\beta(f) d_s}  \prod_{j=1}^{m}(1 - \alpha_j(f)),
\label{eq_19}
\end{equation}    
with $\beta(f)$ a frequency dependent absorbing coefficient \cite{iso}. Furthermore, fixing the sound celerity $c$, room impulse response can be generated, converting each distance $d_s$ in time of arrival. Taking care to convert energy into sound pressure ($p = \sqrt{E}$), finite impulse response can be generated and analyzed using standard metrics (e.g. $T_{30}$, $C_{80}$, $D_{50}$, etc.). For auralization, this room impulse response is convolved with an audio signal in order to listen the acoustical rendering. In particular, this convolution can involve relative position of predominant images sources, in order to realize a spatialized auralization with multichannel or binaural renderers. Finally, to complete acoustic studies with visual analysis, images sources can be projected on the room used for computation to see where are located listened reflections (see last impact on fig. \ref{schema_SI}).

\section{Implementation}
\subsection{Standard algorithm}
As standard principles are introduced, we focus now on the numerical implementation of an acoustic renderer by ray-tracing. Before any acoustic computation, a numerical room has to be modelized with surfaces and materials. In our case, we use the classical representation with mesh composed of flat triangles. Geometrical intersections are computed between rays (represented by oriented lines $(L)$) and mesh elements (represented by pieces of plans $(P)$), using parametric equations :
\begin{eqnarray}
(L) &:& a + \delta \overrightarrow{u}, \quad \delta \in \mathbb{R},  \\
(P) &:& b + \lambda \overrightarrow{v} + \mu \overrightarrow{w}, \quad \lambda, \mu \in \mathbb{R}.
\label{eq_20}
\end{eqnarray}    
Considering $\overrightarrow{v}$ and $\overrightarrow{w}$ driven by two edges of each triangle, the following conditions give pairs (rays;elements)  with uniqueness : 
\begin{itemize}
\item $(0 \leq \lambda \leq 1)$, $(0 \leq \mu \leq 1)$  and $(\lambda+\mu \leq 1)$ to ensure that the intersection is inside the triangle,
\item $\delta > 0$ to respect the propagation direction,
\item $\delta$ minimum not to go through the mesh.
\end{itemize}
Practically, to find these pairs, we can solve directly the underlying linear system or use the Moller-Trumber algorithm \cite{moller}. For $N$ rays and $M$ triangular elements, this process has a quadratic numerical cost (proportional to $NM$), which is critical if both $N$ and $M$ are large (see section \ref{octree}). Once all pairs are found, energy measurement has to be done in order to build images-sources (see section \ref{is}). To this end, the rays are intersected to the measurement sphere $S(x_m, r)$ using its cartesian representation :
\begin{equation}
(x-x_m)^2 - r^2 = 0, \quad \forall x \in  \mathbb{R^3},
\end{equation} 
which leads to a linear numerical cost proportional to $N$.

Finally, a ray is reflected according to equation (\ref{eq_15}) and propagated while its distance travelled verify condition (\ref{eq_13}). This iterative strategy ensure the energy propagation by the elimination of all rays that would be in non-measurable beams. In the particular case of an open-air room, rays which don't encountered surface of the mesh are also eliminated. Once all rays are eliminated, images-sources can be built and post-treated (room impulse response, auralization, etc.).

\subsection{Tree-base acceleration}
\label{octree}
As we have seen, the most critical stage of the standard algorithm is the research of intersections between  rays and triangular elements, leading a priori to a quadratic complexity $O(N M)$. Indeed, each ray has to be tested with each face, for each iteration of the ray-tracing algorithm. For a large number of mesh elements (e.g. $M>10^5$ for the Orange theater) and rays to ensure reasonable accuracy (typically $N>10^6$), the calculation time may be prohibitive. To alleviate this problem, a "Divide and Conquer" approach using binary trees is performed \cite{fft} \cite{matrix}. 

\begin{figure}[t]
\centering
		\includegraphics[width=0.4\linewidth]{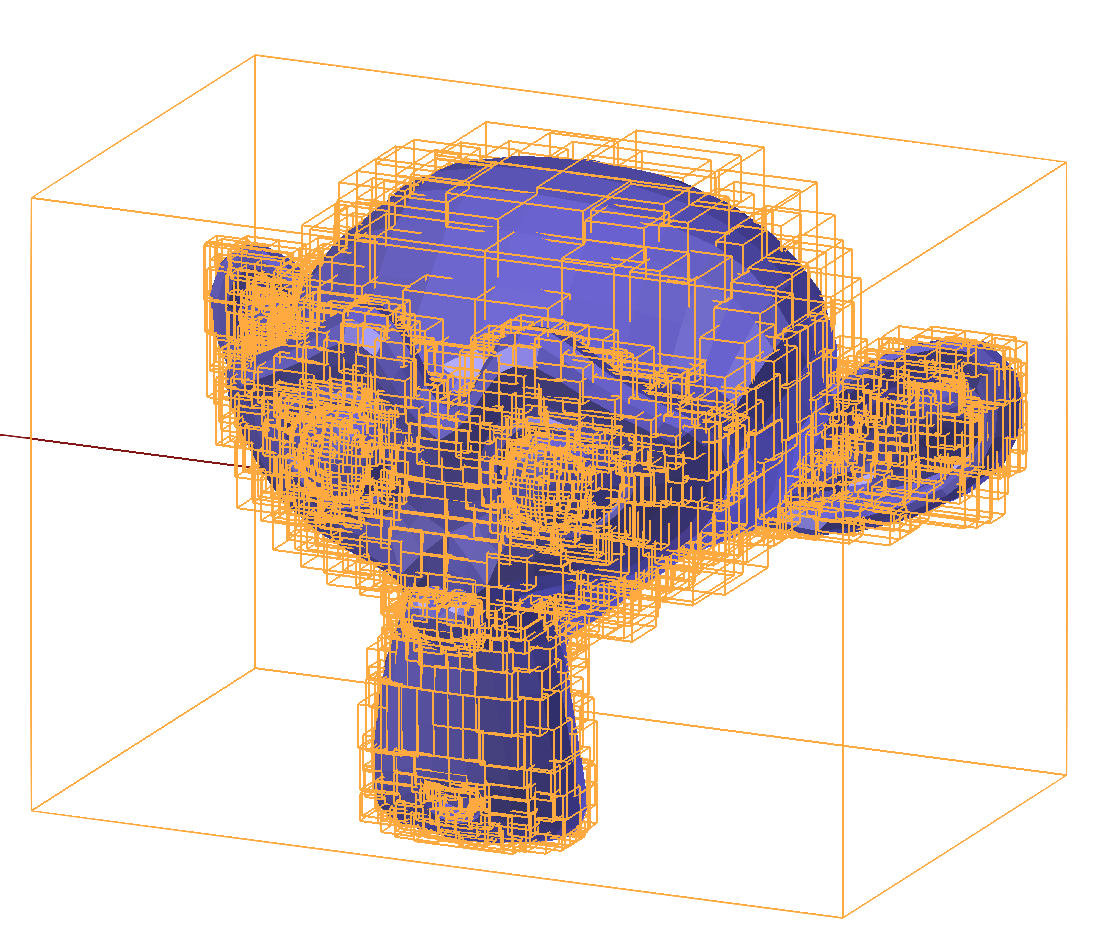}
		\caption{Binary tree leaves of an arbitrary mesh from \textit{Blender}. (computed on \textit{Just4RIR})}
		\label{octreeSuzanne}
\end{figure}

The general principle consists in creating a mother-box, containing all the mesh elements. This mother-box is then subdivided along the largest dimension to create two daughter-boxes, each with the same number of elements (median spatial subdivision). This process is then applied recursively, until a stopping criterion is reached. In our case, we stop when the leaves contain only one element (see fig. \ref{octreeSuzanne}). This hierarchical tree is completely mesh dependent, computed in $O(M\log M)$ operations, and gives a structure which permit to quickly navigate inside the mesh. 

Then, we initialize the ray sorting process starting from the mother-box, containing all rays and elements. Using the first tree-subdivision, we distribute rays inside the two daughter-boxes. This stage is done in $O(N)$ operations by W. Amy \textit{et al.} algorithm \cite{AABB}. Indeed, each box has $N_1$ and $N_2$ rays, such as $N = N_1 + N_2$. Assuming this subdivision is performed recursively to the $p$-level, each box contains $(N_i)$ rays such as: 
\begin{equation}
N = \sum_{i=1}^{2^p} N_i.
\end{equation}  
Then, ray sorting at the $(p+1)$-level also conducts to $O(N)$ operations. To reach the leaves-level, we have to perform $O(N \log M)$ operations, where $\log M$ is close to the depth of the binary tree. At the end, as we have only one element per box, the ray-element intersection only needs $O(N)$ operations. Finally, instead of $O(N M)$ operations, we compute ray-tracing algorithm in~: 
\begin{equation}
O(M\log M) + O(N \log M) + O(N),
\end{equation}  
witch is a near-linear complexity. More over, if binary tree is precomputed, each iteration of ray-tracing just stands for :
\begin{equation}
O(N \log M) + O(N).
\end{equation}
 
\begin{figure}[t]
\centering
	\includegraphics[width=0.8\linewidth]{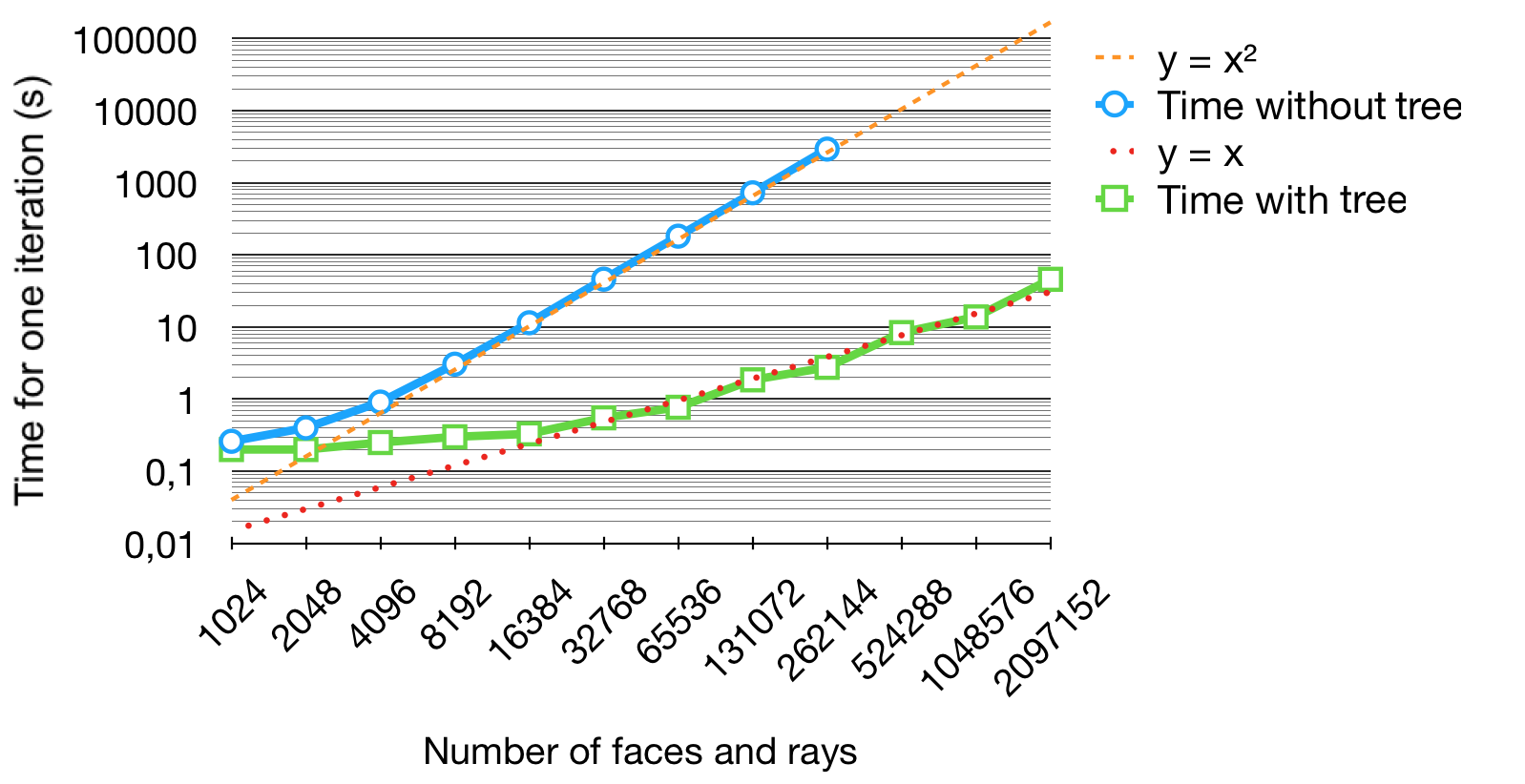}
	\caption{Computation time for one iteration of ray-tracing in function of the number of face and rays, such as $N = M$ (log~scale). An omnidirectional source is located at the center of a mesh of a unitary tetrahedral. (computed on \textit{Just4RIR})}
	\label{times}
\end{figure}

To evaluate numerically complexities with or without binary tree acceleration, we measure the computation time of one iteration by increasing the number of rays and the number of faces in the mesh ($N=M$). As we can see in figure \ref{times}, the complexity of the algorithm is therefore quite linear by using tree-based method. This allows to treat large meshes with millions of rays, by maintaining a reasonable computation time. In particular, we can see in the table \ref{tabComplexite} that for 250~000 rays and faces the computation time is divided by a factor of a thousand compared to the classical method. It is important to precise that no parallelization strategy has yet been employed, all results are obtained in single core computation on a standard laptop (2.7 GHz core and 8 Go ram).

\begin{table}[t]
\centering
	\begin{tabular}{| c | c | c |}
		\hline
		Number of faces and rays & Time \textbf{without} tree (s) & Time \textbf{with} tree (s)\\
		  \hline
		  \hline
		   $2^{10}$ (=1~024) & 0,26 &	0,2 \\
		   \hline
		$2^{11}$ (=2~048)  & 0,4	& 0,2 \\
		   \hline
		$2^{12}$ (=4~096) & 0,91	& 0,25\\
		   \hline
		$2^{13}$ (=8~192) & 3,05 &	0,3\\
		   \hline
		$2^{14}$ (=16~384) & 11,44	&0,33\\
		   \hline
		$2^{15}$ (=32~768) & 46,02	&0,55 \\
		     \hline
		    $2^{16}$ (=65~536) & 181,61	& 0,77\\
		   \hline
		$2^{17}$ (=131~072) & 725,17	& 1,85\\
		\hline
		$2^{18}$ (=262~144) & 2927,9 & 2,76 \\
		\hline
		$2^{19}$ (=524~288) & X & 8,36 \\
		\hline
		$2^{20}$ (=1~048~576) & X & 13,78 \\
		\hline
	 \end{tabular}
	\caption{Computation time of fig. \ref{times} on a standard laptop (2.7 GHz core and 8 Go ram).}
	\label{tabComplexite}
\end{table}

\section{Numerical validation}
To evaluate and validate methods and algorithms, several non-regression tests have been implemented. In this study, we focus only on two examples, that we hope will be the most significant. For more details, readers are referred to R. Gueguen's PhD manuscript \cite{theseRobin} or \textit{Gypsilab} examples in source code \cite{githubGypsi}.   

\subsection{Energy conservation by reflecting sphere}
\begin{figure}[t]
\centering
		\includegraphics[width=0.6\linewidth]{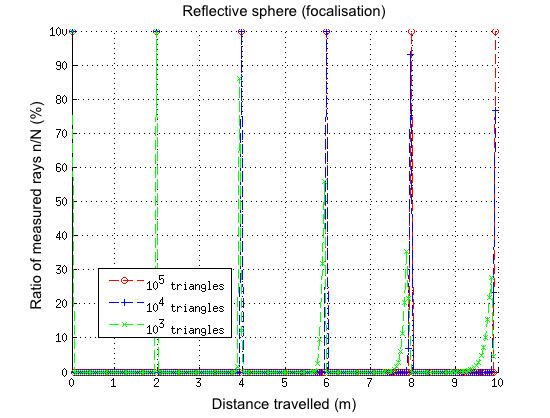}
		\caption{Energy measured according to equation \ref{eq_7} for a 100\% reflecting unit sphere, with punctual source and measure located at the center (in percent). This result is given after six iterations of the tree-based algorithm, and we note as expected a two meters distance between pulses. $10^5$ rays have been used. (computed on \textit{Gypsilab})}
		\label{test2RIR}
\end{figure}

Firstly, we want to prove that the statistical approach combined with the fast algorithm is able to conserve the acoustical energy (eq. \ref{eq_2}). To do so, we consider an academic problem, where the room is a $100\%$ reflecting unit sphere, containing punctual source and measure located at the center. If we consider only specular reflections with no air-absorption, it is expected that energy measurement takes out a Dirac comb, two meters spaced.
Numerically, as we discretize the sphere with flat triangles, this focusing property clearly depend of the mesh refinement. As shown on figure \ref{test2RIR}, numerical diffusion appears in function of the distance.  At the end, only the $10^5$ elements mesh is able to maintain a pulsed energy in the range given, and this computation needs fast methods. Practically, this problem is representative of curved surfaces in rooms, which need high refinement to ensure a good propagation of the acoustic energy.

\subsection{Analytical comparison by shoe box model}
Secondly, numerical computation is compared to the analytical shoe box model, introduced by S. McGovern \textit{et al.} \cite{mcgovern}. This model gives image-sources positions at any reflection order, as a function of the punctual source and measure located inside the box. Originally, the energy associated to each image-source is given by :
\begin{itemize}
\item Quadratic decay $1/d^2$, where $d$ stands for image-sources distances to measurement point,  
\item Absorption coefficients $\alpha$ for each wall, with no frequency dependance.
\end{itemize}
To use this model for our validation process, we add to McGovern solutions a frequency dependance for wall absorption, and the atmospheric impact in function of the distance. This extended model gives the energy as equation (\ref{eq_19}), but the statistical counting is there replaced by the distance quadratic decrease.

\begin{figure}[t]
\centering
	\begin{subfigure}{0.47\textwidth}
		\includegraphics[width=\linewidth]{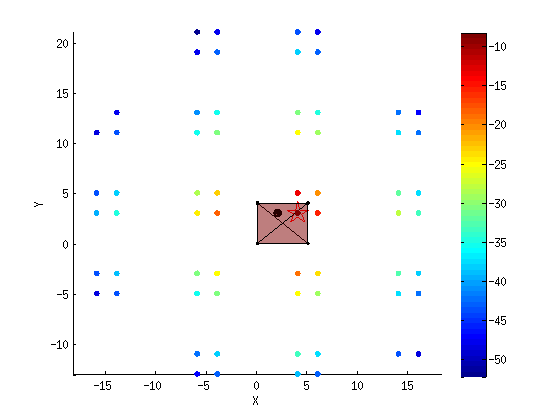}
		\caption{image-sources in the plan z = 1.7 m and associated energy (dB). Analytical and numerical solutions are the same, up to machine precision.}
		\label{sourcesImages}
	\end{subfigure}
	\quad
	\begin{subfigure}{0.47\textwidth}
		\includegraphics[width=\linewidth]{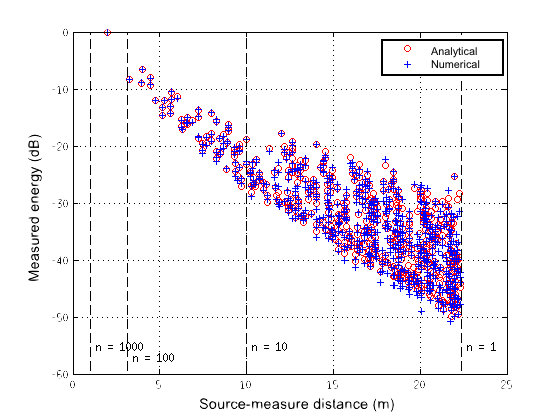}
		\caption{Energy impulse response in dB. Even if accuracy decreases with the distance, far solutions remain acceptable to describe the diffuse field. }
		\label{boite}
	\end{subfigure}
	\caption{Numerical solutions from ray-tracing inside a shoes box model. Comparisons with analytical solutions. (computed on \textit{Gypsilab})}
\end{figure}

To compare analytical extended model and numerical ray-tracing computation, we use a simple triangular mesh of $M=12$ triangular elements building a $[5,4,3]$ m box. A randomly location is given for source ($x_s = [4,2,1.7]$ m) and an arbitrary measurement sphere is fixed ($x_s = [2,2,1.7]$ m and $r=0.2$ m). Wall absorption coefficients are chosen randomly in the \textit{Odeon} database \cite{odeon}, and atmospheric model is used following the norm ISO-9613-1\cite{iso}. $N = 10^ 5$ rays were used and iterations were stopped according to criterion (\ref{eq_13}). As a result, figure \ref{sourcesImages} shows image-sources computed which fit exactly to analytic positions (machine accuracy). As tri-dimensional representation is really confusing, we only represent images sources in the plan $z = 1.7$m. Furthermore, energy value is given in decibel on the color bar, in order to visually see the distance decrease. More precisely, figure \ref{boite} shows a comparison of the energy in decibel, in function of the distance of source images. As expected, a good matching is reached for direct sound and early reflections and diffuse field seems well approached. 
 
\section{Application to Orange theater}

The fast ray-tracing algorithm allows acoustical computation for rooms as complex as the ancient theater of Orange. It is of particular importance since using a virtual model, archeologists can explore many architectural hypotheses. In particular, acoustical analysis allows to understand behaviors like : the influence of the position of spectators in bleachers, the shape of the roof, the materials of the \textit{orchestra}, etc. Moreover, it is interesting to study where reflections come from, and more generally how the theater responds to acoustical sources in various locations.     

\begin{figure}[t]
\centering
		\includegraphics[width=0.9\linewidth]{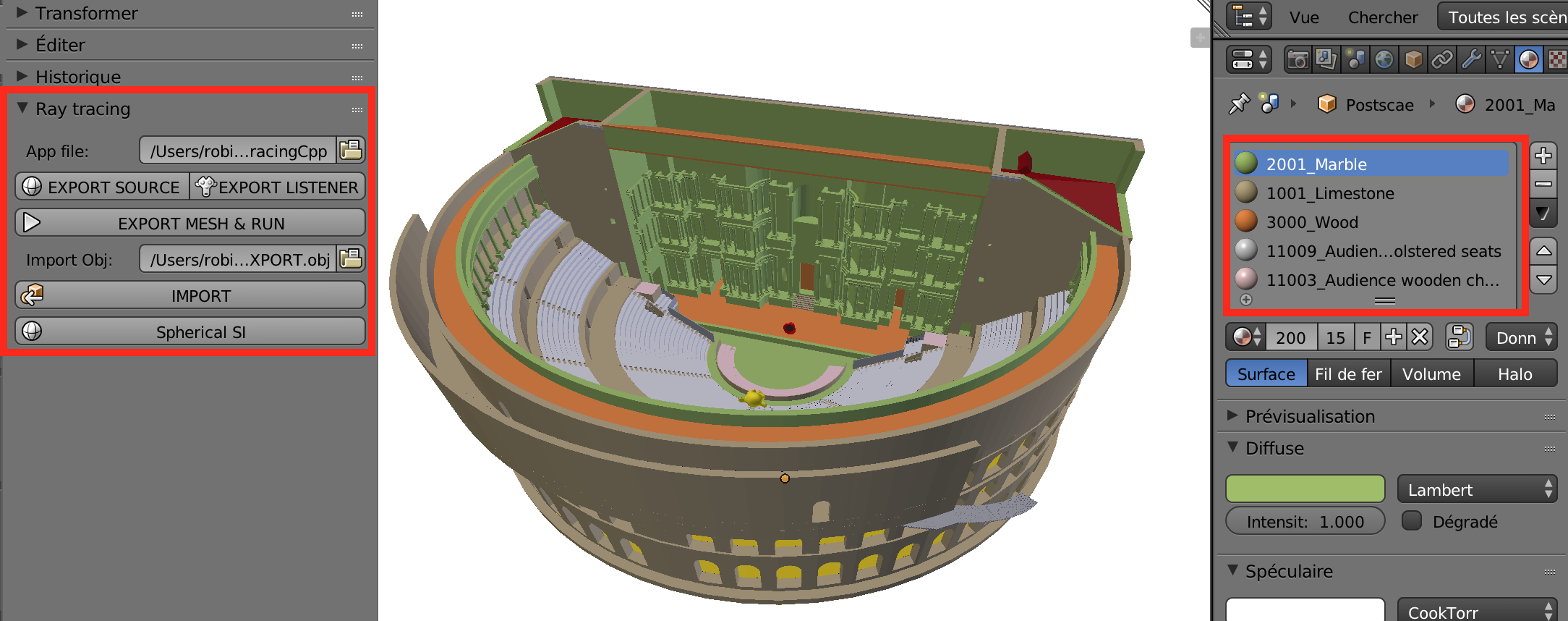}
		\caption{Modelization of restituted theater of Orange \cite{theseRobin} in \textit{Blender} showing the ray-tracing add-on on the left and the material interface on the right.}
		\label{soft}
\end{figure}

First of all, a restituted version of the Orange theater has been realized on the software \textit{Blender} (fig. \ref{soft}), according to the archeological surveys performed by l'Institut de Recherche sur l'Architecture Antique (IRAA) \cite{orangeTxt}. As this mesh just needs to fill the geometry, there is no need of Delaunay properties, but the architecture complexity however conducts to a 436~000 faces triangulation (see fig. \ref{maillage}). In particular, this virtual resitution is mainly composed by :
\begin{itemize}
 \item \textit{Postscaenium} (stage wall) partially ornamented with two basilicas on either side,
 \item \textit{Pulpitum} (stage),
 \item \textit{Orchestra},
 \item \textit{Cavea} (bleachers) with \textit{Porticus} at the top (column gallery),
  \item Various covers (stage roof, \textit{velum}, etc.).
\end{itemize}
Several materials are assigned to each part of the theater, in order to define specific absorption coefficients taken from \textit{Odeon} database \cite{odeon} (see fig. \ref{soft}). The ray-tracing solver is then used to compute the spatial impulse response. To reach the reverberation time close to -60dB ($RT_{60}$), we fix one million rays and a 2m-radius measurement sphere. The result is obtained in few minutes on a standard laptop (2.7 GHz core and 8 Go ram). At the end, all sources-images positions are generated as well as the associated multi-band acoustical energy (i.e. couple $(x_s;E_s(f))_{s \in [1, N_s]}$, see section \ref{is}). Eventually, multi-outputs are analyzed in order to study the acoustical behavior of the theater.

\begin{figure}[t]
\centering
	\begin{subfigure}{0.6\textwidth}
		\includegraphics[width=\linewidth]{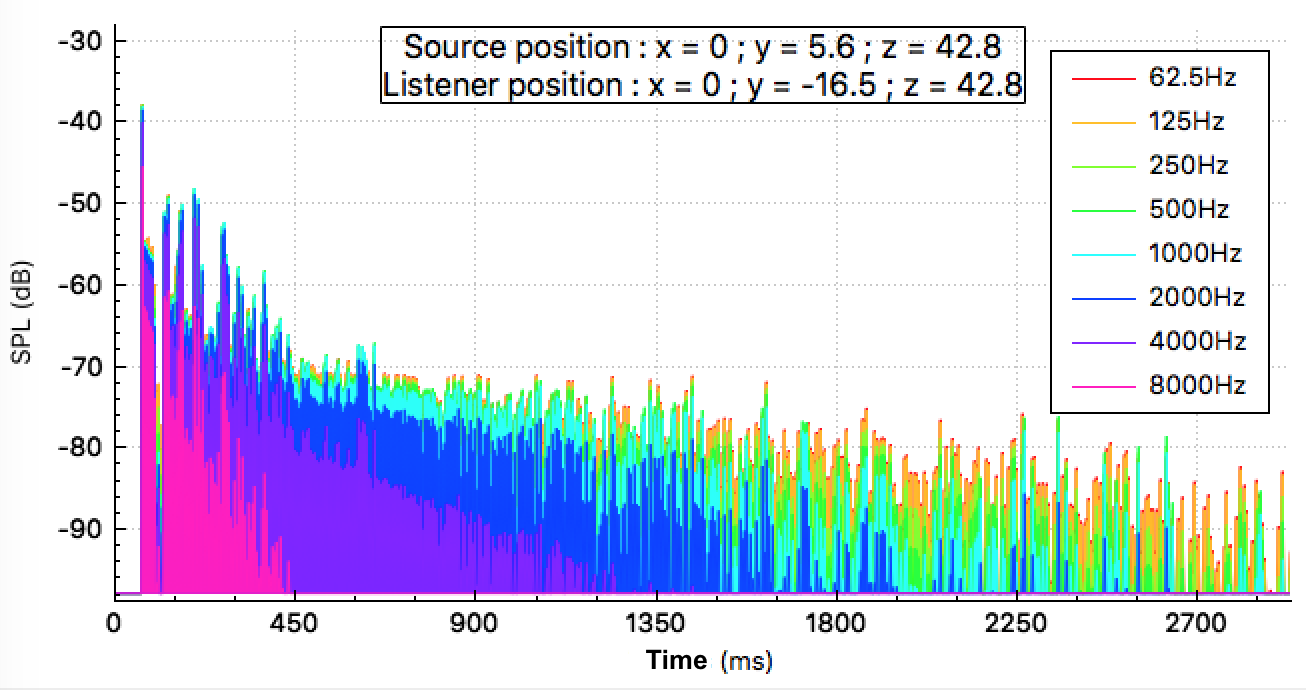}
		\caption{Impulse response in dB.}
		\label{rirTheatre20}
	\end{subfigure}
	\begin{subfigure}{0.37\textwidth}
\includegraphics[width=\linewidth]{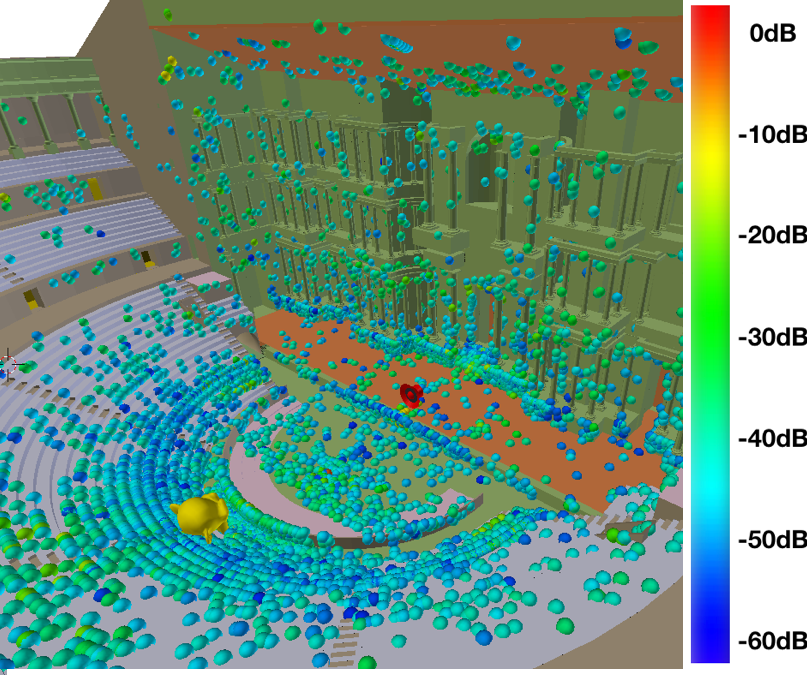}
		\caption{Images-sources projected on the mesh.}
		\label{isTheatre20}
	\end{subfigure}
	\caption{Ray-tracing outputs for one million rays and a 2m-radius sphere (listener) on restituted theater of Orange.  (computed on \textit{Just4RIR})}
	\label{SI60dB}
\end{figure}

\begin{figure}[t]
\centering
	\begin{subfigure}{0.63\textwidth}
		\includegraphics[width=\linewidth]{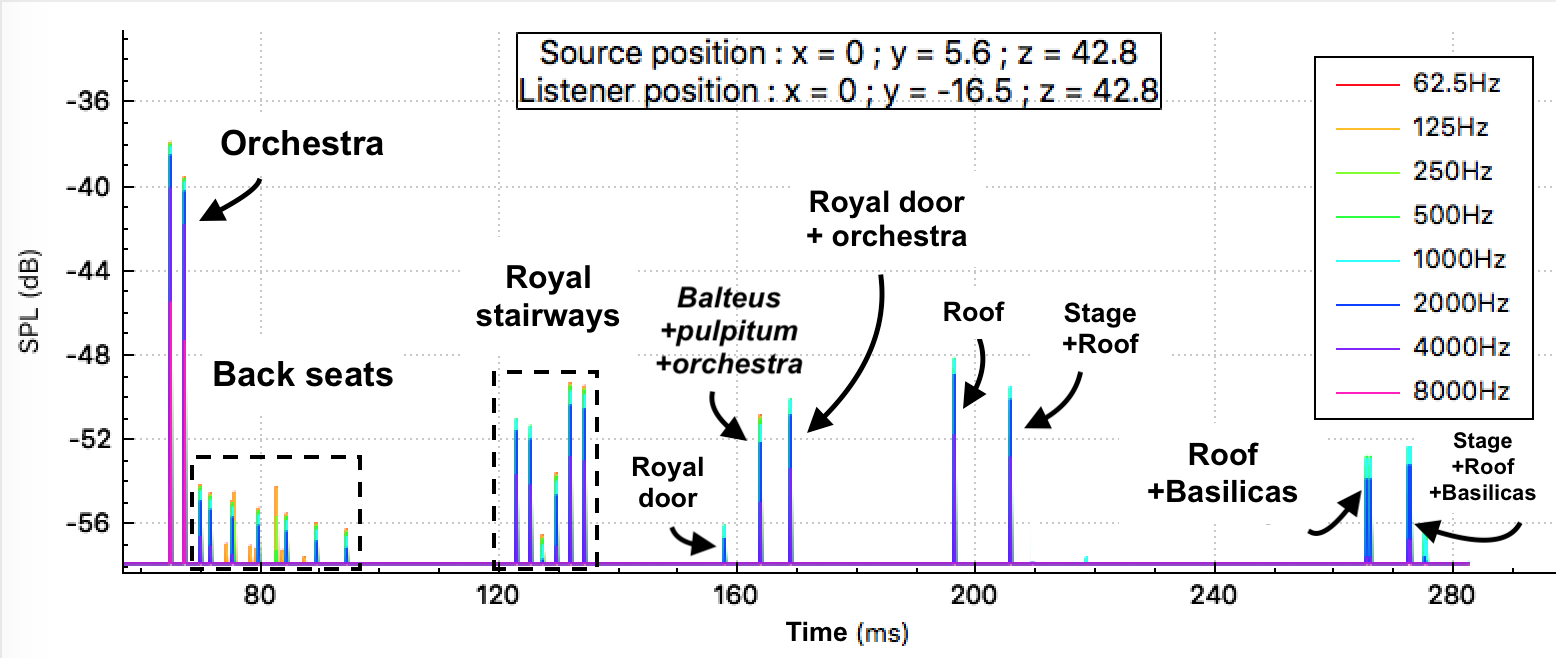}
		\caption{Impulse response of the first reflections over 20dB range.}	
		\label{RIR20dB}
	\end{subfigure}
	\begin{subfigure}{0.36\textwidth}
		\includegraphics[width=\linewidth]{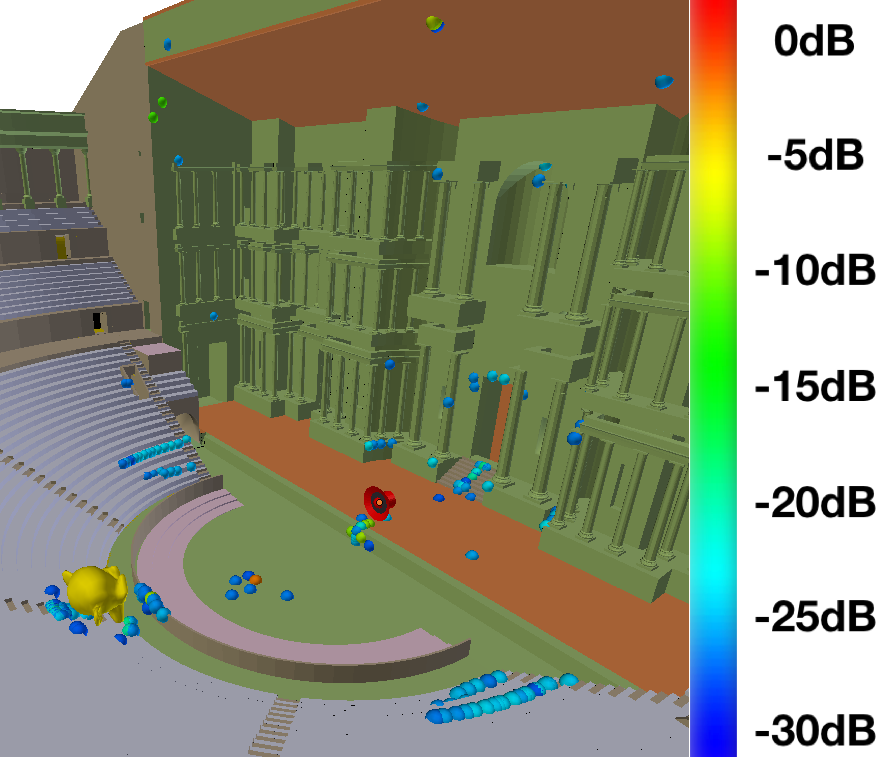}
		\caption{Images-sources of the first reflections over 30dB range, projected on the mesh.}
		\label{SI30dB}
	\end{subfigure}
	\caption{Zooming on the first reflections of fig. \ref{SI60dB}. (computed on \textit{Just4RIR})}
	\label{SItheater}
\end{figure}

The following results are obtained with a source located at the front stage, 1,60 m above the floor (this correspond to the position of the mouth of an average actor). The listener is on the same axis in the bleachers. On figure \ref{rirTheatre20}, we see the multi-band impulse response of the theater until the maximum distance determined by equation (\ref{eq_13}). Primary reflections appear on the first 400 ms and diffuse field decreases according to the frequency. This difference is due to atmospheric absorption and materials properties and only frequencies above 1kHz reach $RT_{60}$. Furthermore, the projected images-sources represented on figure \ref{isTheatre20} illustrate a spatial diffusion. Indeed, even if some areas carry a lot of images-sources, reflections seem to surround the listener. For a better understanding, we zoom on the top of the signal in figure \ref{SItheater}. We can notice high contribution of the \textit{orchestra}, the wall-stage, the stage and the roof, as F. Canac demonstrated in the 60' \cite{canac}. 

To finish, perceptive factors were computed from the energy impulse responses, and given in table \ref{tab_rindel}. Results correspond to a room adapted for musical playback (e.g. clarty $C_{80}$ and reverberation time $T_{30}$), more than a speech transmission \cite{acoustique}. These results are confirmed by an equivalent simulation in the \textit{Odeon} software (commercial license), using the same mesh and parameters. 

\begin{table}[h]
\centering
 \begin{tabular}{| *{3}{c|}} 
 \hline 
 Perceptive factors & \textit{Just4RIR} & \textit{Odeon} \\ 
 \hline 
 \hline 
  EDT (s)& 1.85& 2.1 \\ 
 \hline 
$T_{30}$ (s)& 3.59&  2.73\\ 
 \hline 
SPL (dB) &-32 & -32.4\\ 
 \hline 
$C_{80}$ (dB)& 1.12&1.5  \\ 
 \hline 
$D_{50}$ (\%)&47 & 52 \\ 
 \hline 
$T_s$ (ms)&117 & 105 \\ 
 \hline 
\end{tabular} 
 \caption{Perceptive factors obtained for the Orange theater, comparison between \textit{Just4RIR} and \textit{Odeon} software \cite{odeon}. The measure is done on the 500-1000Hz band.}
 \label{tab_rindel} 
 \end{table}

\section{Conclusions}

In this paper, a full-chain engineering process is given, leading by an acoustical study of an imposing ancient monument. From archaeological needs, a fine mesh of the monument was constructed, associated to a complete room acoustic application suite, both for {\sc Matlab} and \textit{Blender}. The high complexity provided by this type of architecture and its ornaments leads to approximate calculation methods. Indeed, by only simulating specular reflections and wall absorption, energy propagation and measurement can be simulated by beams, carried by ray-tracing. From this representation basis, it is possible to generate a multi-band impulse response, while respecting the laws of high-frequency acoustics. Moreover, a fast algorithm with a near-linear complexity has been implemented, allowing users to quickly evaluate architectural assumptions, modifying their meshes regardless of the number of elements. At the end, various post-treatments have been added, as the Room Impulse Response generation, the source-image visualization, the classical perceptive factors and an auralization process.

Even if current versions of proposed softwares (\textit{Gypsilab} \cite{githubGypsi} and \textit{Just4RIR}) are complete enough to be used for various studies, there are many opportunities of improvement. First of all, as image-sources positions are known, a spatial audio renderer should be added to improve current auralization tool (e.g. binaural or multichannel, eventually with trackers). Secondly, as virtual reality is becoming more and more important in today's applications, we could consider moving the listener in real time and thus, allow a complete virtual tour of the building. Finally, as ray-tracing modelization is an high-frequency approximation of waves phenomena, diffraction effects should be added in order to get a better fit with the physical phenomena.

\section*{Acknowledgments}
The authors are particularly indebted to Fran\c{c}ois Alouges, Titien Bartette, Pascal Frey and Emmanuelle Rosso for the help they all provided at the different stages of this project. Thanks also to Jean-Dominique Polack for advices on architectural acoustics and Martin Lesellier for various contributions. This work is part of R. Gueguen PhD thesis founded by Sorbonne Universit\'e.

\bibliography{Biblio}

\end{document}